\documentclass[a4paper,11pt,oneside]{article}

\usepackage{geometry} 
\usepackage{amsmath}
\usepackage{amssymb}
\usepackage{color}
\usepackage{graphicx}
\usepackage{booktabs}
\usepackage[small,bf]{caption}
\newcommand{\fracwithdelims}[4]{\Bigl#1 \frac{#3}{#4} \Bigr#2}

\newcommand{\eq}[1]{eq.~(\ref{eq:#1})}

\newcommand{\nohyphens}%
       {\hyphenpenalty=10000\exhyphenpenalty=10000\relax}

\DeclareMathOperator{\diag}{Diag}

\def\pM{\ensuremath{\genfrac{}{}{0pt}{1}{+}{\scriptstyle(\kern-1pt-\kern-1pt)}}}

\def\gtap{\mathrel{ \rlap{\raise 0.511ex \hbox{$>$}}{\lower 0.511ex
   \hbox{$\sim$}}}} 
\def\ltap{\mathrel{ \rlap{\raise 0.511ex
   \hbox{$<$}}{\lower 0.511ex \hbox{$\sim$}}}}

\allowdisplaybreaks[1]

\newlength{\myem}
\newcommand{\sep}[1]{#1}
\newcounter{mysubequation}[equation]
\renewcommand{\themysubequation}{\alph{mysubequation}}
\newcommand{\mytag}{\stepcounter{mysubequation}%
\tag{\theequation\protect\sep{\themysubequation}}}
\newcommand{\globallabel}[1]{\refstepcounter{equation}\label{#1}}

\newcommand{\SISSA}{SISSA/ISAS and INFN, I--34136 Trieste, Italy}

\newcommand{\preprintnumber}{%
SISSA  40/2011/EP}
 
\newcommand{\titletext}
 {Sizeable $\boldsymbol{\theta_{13}}$ from the Charged Lepton Sector in SU(5),
 (Tri-)Bimaximal Neutrino Mixing and Dirac CP Violation}
\newcommand{\authortext}{
\large
David Marzocca$^{\, a}$,
Serguey T. Petcov$^{\, a, b, c}$,
Andrea Romanino$^{\, a}$,
Martin Spinrath$^{\, a}$
\medskip\\\em\normalsize 
$\mbox{}^a$ \SISSA
\\[0.1\baselineskip] 
$\mbox{}^b$ IPMU, University of Tokyo, Tokyo, Japan
\\[0.1\baselineskip] 
$\mbox{}^c$ Institute of Nuclear Research and Nuclear Energy, Bulgarian Academy of Sciences, 1784 Sofia, Bulgaria.}

\title{
\normalsize
\hspace*{\fill}
\begin{tabular}[t]{l}\preprintnumber\end{tabular}
\vspace{3\baselineskip}\\\Large\bfseries\titletext\bigskip}
\author{\begin{minipage}[t]{0.8\textwidth}
\normalsize\centering\authortext
\end{minipage}}
\date{}

\begin{document}

\maketitle

\begin{abstract}
\noindent
The recent results from T2K and MINOS experiments 
point towards a relatively large value of the reactor angle 
$\theta_{13}$ in the lepton sector. In this paper we show how
a large $\theta_{13}$ can arise from the charged lepton 
sector alone in the context of an SU(5) GUT. In such a scenario (tri-)bimaximal mixing in the neutrino sector is  still a viable possibility. We also analyse
the general implications of the considered scenario
for the searches of CP violation in neutrino oscillations.
\end{abstract}

\section{Introduction}

 Recently the T2K collaboration reported \cite{Abe:2011sj} 
evidence at $2.5\sigma$ for a non-zero   
value of the reactor angle $\theta_{13}$ 
in the neutrino mixing matrix. 
Quantitatively, it was found that 
$0.03(0.04) < \sin^2 2\theta_{13} < 0.28(0.34)$ at 90\% C.L.\
for $\Delta m^2_{32} =  \pM2.4\times 10^{-3}~{\rm eV^2}$, 
$\sin^22\theta_{23} = 1$ and $\delta = 0$, $\Delta m^2_{32}$ and
$\theta_{23}$ being the atmospheric neutrino 
mass squared difference and mixing angle, 
and $\delta$ being the Dirac 
CP violating phase in the neutrino mixing matrix 
(see, e.g.\ \cite{PDG10}). Under the same conditions 
the best fit value of $\theta_{13}$ obtained 
from the data analysis is relatively large: 
$\sin^22\theta_{13}= 0.11~(0.14)$.
Subsequently the MINOS collaboration also reported   
evidence for a relatively large value of $\theta_{13}$, 
although with a smaller statistical significance \cite{MINOS240611}.
A global analysis of the neutrino oscillation data, including 
the data from the T2K and MINOS experiments, performed 
in \cite{Fogli:2011qn} showed  that actually 
$\sin\theta_{13}\neq 0$ at $\geq 3\sigma$.
The authors of \cite{Fogli:2011qn} find:
\begin{equation}
\sin^2\theta_{13} = 0.021~(0.025) \pm 0.007\,,
\label{th13Fogli}
\end{equation}
%
using the ``old'' (``new'') fluxes of reactor 
$\bar{\nu}_e$ in the analysis. Moreover, it was found in the same global 
analysis that $\cos\delta = -1$ (and
$\sin\theta_{13}\cos\delta = - 0.14$) 
is clearly favored by the data over 
$\cos\delta = + 1$ 
(and $\sin\theta_{13}\cos\delta = + 0.14$~
\footnote{This result can have important implications for the
``flavoured'' leptogenesis scenario of
generation of the baryon asymmetry of the Universe
\cite{EMSTP08}.} 
).

The T2K and MINOS results will be tested in the upcoming 
reactor neutrino experiments Double Chooz \cite{DCHOOZ}, 
Daya Bay \cite{DayaB} and RENO \cite{RENO}.
If confirmed, they will have far reaching implications 
for the program of future research in neutrino physics. 
A relatively large value of $\theta_{13}$ 
opens up the possibilities, in particular,  
i) for searching for CP violation effects in 
neutrino oscillations 
experiments with high intensity accelerator 
neutrino beams (like T2K, NO$\nu$A, etc.);
ii) for determining the sign of 
$\Delta m^2_{32}$, and thus the type 
of neutrino mass spectrum, which can be 
normally or invertedly ordered (see, e.g.\ \cite{PDG10}), 
in the long baseline neutrino oscillation experiments 
at accelerators (NO$\nu$A, etc.), 
in the experiments studying the oscillations 
of atmospheric neutrinos (see, e.g.\ \cite{JBSP203}), 
as well as in experiments with reactor antineutrinos
\cite{ReactNuHiera}.
A value of $\sin\theta_{13} \gtap 0.09$ 
is a necessary condition for a successful ``flavoured''
leptogenesis when the CP violation required for 
the generation of the matter-antimatter asymmetry of the
Universe is provided entirely by the Dirac CP violating
phase in the neutrino mixing matrix \cite{PPRio106}.

In the present article we investigate the possibilities 
to obtain the relatively large value of $\theta_{13}$, 
suggested by the T2K, MINOS and the current global neutrino 
oscillation data, in
the case 
when the diagonalization 
of the neutrino mass matrix gives a negligible 
contribution to $\theta_{13}$.
Parametrically the latter 
means a contribution smaller than $\mathcal{O}(\lambda^2)$, where $\lambda=0.23$ 
is the Cabibbo angle expansion parameter.
The diagonalization of the neutrino mass matrix leads 
to a negligible $\theta_{13}$,
for example, in models with $L_e - L_{\mu} - L_{\tau}$ 
flavour symmetry \cite{SPPD82} and in
most models accounting for bimaximal (BM) \cite{bima} 
or tri-bimaximal (TBM) \cite{tri} neutrino mixing.
For our purposes, this is equivalent to setting 
the neutrino mass matrix contribution to $\theta_{13}$ to zero, 
which we will do in our further analysis. 
The $\theta_{13}$ mixing angle 
can
then be due to the contribution to the 
neutrino mixing coming from the diagonalization of 
the charged lepton mass matrix 
(see, e.g.\ \cite{FPR04,Romanino:2004ww,HPR07} 
and the references quoted therein 
\footnote{For alternative possibilities leading to a 
relatively large $\theta_{13}$, see \cite{GPRR09} as well as, e.g.\ \cite{LargeTh13}.} 
).

In particular, we will consider the possibility 
of a ``large'' $\theta_{13}$ originating 
from the charged lepton mass matrix
in the context of  SU(5) supersymmetric unification. 
The motivation for such a choice are strong and well known. 
In our case, this choice will also allow to get a 
handle on the expected contribution to $\theta_{13}$ and 
on the structure of the charged lepton 
Yukawa coupling matrix. This is because the SU(5) symmetry 
allows to relate the latter and the down quark one, 
on which further experimental information is available.

In minimal SU(5) models, the down-quark and 
transposed charged lepton mass matrices are equal. 
On the other hand, depending on the (renormalizable 
or non-renormalizable) SU(5) 
operator(s) giving rise to a given Yukawa coupling matrix entry, the 
relation may involve Clebsch-Gordan (CG) coefficients. Such 
coefficients need to be invoked in order to fix the SU(5) 
predictions for the $m_b/m_\tau$ and $m_s/m_\mu$ mass ratios
(unless specific values of $\tan\beta$, peculiar patterns of SUSY breaking 
parameters, or highly asymmetric mass textures are 
employed~\cite{bTauUnification}), as shown by the authors 
of~\cite{Antusch:2008tf, Antusch:2009gu} after having identified the 
possible GUT scale mass ratios in a bottom-up approach. All the 
possible CG coefficients in SU(5) and the Pati--Salam group 
entering such relations arising from dimension five and 
some dimension six operators are listed in \cite{Antusch:2009gu}.
Such coefficients not only can give rise to new mass ratios, 
but can also affect 
the prediction for $\theta_{13}$~\cite{Antusch:2009gu}. 
In this paper, we discuss 
possible combinations of CG coefficients, 
which can give a ``large'' $\theta_{13}$ 
even if there is no contribution to $\theta_{13}$ 
from the neutrino mass matrix at all.

The paper is organised as follows. In Section 2 we 
describe the general set up, 
the determination of the standard neutrino parameters 
under the assumptions we make, 
the numerical procedure we used to determine the allowed CG coefficients and 
the predictions for $\theta_{13}$. The results are illustrated in Section 3. 
In Section 4 we consider the special case in which the neutrino 
contribution to the PMNS matrix 
is of the BM or TBM form, in which interesting implications for the leptonic 
(Dirac) CP-violation phase can be drawn. In Section 5 we summarize.

\section{General Setup and Procedure}

\subsection{The Setup}

We start with some definitions and a summary of the assumptions we 
make regarding the structure of the mixing matrices. 
Thereby we follow closely the discussion in 
\cite{Romanino:2004ww,FPR04,HPR07}. 
The PMNS lepton mixing matrix $U$ is given by 
\begin{equation}
\label{eq:U}
U = U_e U^\dagger_\nu,\quad \text{with $U_e$, $U_\nu$ defined by} \quad 
\begin{aligned}
m_E = U^T_{e^c} m^\text{diag}_E U_e \\
m_\nu = U^T_\nu m^\text{diag}_\nu U_\nu 
\end{aligned} ,
\end{equation}
%
where $m_E$ and $m_\nu$ are the charged lepton and light 
neutrino mass matrices, respectively, and $m^\text{diag}_E, 
m^\text{diag}_\nu$ are diagonal and positive. 
The PMNS matrix can be parameterized in the standard way as 
$U = R_{23}(\theta_{23}) R_{13}(\theta_{13},\delta) 
R_{12}(\theta_{12}) Q$, 
where $Q$ is a diagonal phase matrix containing 
the two physical Majorana CP violation phases 
\cite{BHP80,SchVall80}.

Barring correlations among the entries of $m_E$, 
the hierarchy of the charged lepton masses 
translates into the possibility of diagonalizing 
$m_E$ perturbatively through subsequent 
$2\times 2$ unitary rotations in the following 
order: $U_e = U^e_{12} U^e_{13} U^e_{23}$, 
where $U^e_{ij}$ is a unitary 
rotation in the $ij$ block. 
The 13 rotation is negligible for our purposes and 
will be set to zero, thus  $U^e_{13} = \mathbf{1}$.

The assumption of zero neutrino mass matrix contribution to 
$\theta_{13}$ can be rephrased as $\theta_{13} = 0$ in 
the limit of $U_e = \mathbf{1}$, or equivalently as 
$U^\dagger_\nu = U|_{U^e = \mathbf{1}} = 
R_{23}(\theta^\nu_{23}) R_{12}(\theta^\nu_{12})$ up to phase matrices. 
Note that since $U_{23}^e$ 
does not contribute to $\theta_{13}$, the assumption 
can be further rephrased as 
\begin{multline}
\label{eq:assumption}
U = R_{12}(\theta^e_{12}) \Phi R_{23}(\hat \theta_{23}) R_{12}(\hat \theta_{12})Q
= \\
\begin{pmatrix}
\hat c_{12} c^e_{12} - \hat s_{12} \hat c_{23} s^e_{12} e^{i\phi} &
\hat s_{12} c^e_{12} + \hat c_{12} \hat  c_{23} s^e_{12} e^{i\phi} & 
\hat s_{23} s^e_{12} e^{i\phi} \\
-\hat c_{12} s^e_{12} - \hat s_{12} \hat c_{23} c^e_{12} e^{i\phi} &
\hat c_{12} \hat c_{23} c^e_{12} e^{i\phi} -\hat s_{12} s^e_{12} &
\hat s_{23} c^e_{12} e^{i\phi} \\
\hat s_{12} \hat s_{23} &
-\hat c_{12} \hat s_{23} &
\hat c_{23}
\end{pmatrix}~Q
.
\end{multline}
Here $\Phi = \diag(1,e^{i\phi},1)$, 
where  $\phi$ is a CP-violating phase~ \cite{FPR04}. 
The angles $\theta^e_{12}$, $\hat\theta_{12}$, $\hat\theta_{23}$ and the 
phase $\phi$  
in the expression (\ref{eq:assumption}) for $U$ can be 
arranged to lie in the intervals $[0,\pi/2]$ and $[0,2\pi]$,
respectively.

By commuting with the $\hat \theta_{23}$ one, 
the $\theta_{12}^e$-rotation has induced a 
non vanishing $\theta_{13}$ given by
\begin{equation}
\label{eq:t13}
\sin\theta_{13} = |U_{e3}| = 
\frac{\sin\theta_{12}^e\tan\theta_{23}}{\sqrt{1+\tan^2\theta_{23} -\sin^2\theta^e_{12}}} \approx \sin\theta_{12}^e \sin\theta_{23} .
\end{equation}
Note that we have traded here $\hat{\theta}_{23}$ 
with $\theta_{23}$ in the standard parameterization. 
Note also that the corrections to 
$\hat \theta_{23}$ and $\hat \theta_{12}$ induced by
$\theta^e_{12}$ are given by 
\globallabel{eq:corrections}
\begin{align}
\tan\theta_{23} &= \cos\theta^e_{12} \tan\hat\theta_{23} \mytag \\
\tan\theta_{12} &= \tan\hat\theta_{12} \left|
\frac{1+\cos \hat \theta_{23} \tan\theta^e_{12} e^{i\phi}}
{1-\tan\hat\theta_{12}\cos\hat \theta_{23} \tan\theta^e_{12}e^{i\phi}}
\right| . \mytag 
\end{align}

In the case of 3-$\nu$ mixing under discussion, 
the magnitude of the CP violation effects in neutrino 
oscillations is determined \cite{PKSP3nu88} by
the rephasing invariant associated  with 
the Dirac CP violating phase $\delta$:
\begin{equation}
J_{\rm CP} = {\rm Im}\left\{ 
U_{e1}^\ast \, U_{\mu 3}^\ast \, U_{e 3} \,
U_{\mu 1} \right\}. 
\label{JCP}
\end{equation}
%
The rephasing invariant $J_{\rm CP}$, 
as is well known, 
is a directly observable quantity. 
It is analogous to the 
rephasing invariant associated with the 
Dirac phase in the Cabibbo-Kobayashi-Maskawa
quark mixing matrix, introduced 
in \cite{CJ85}. 
In the standard parametrisation of the PMNS matrix we find
\begin{equation}
J_{\rm CP} = \frac{1}{8}\,\sin2\theta_{23}\,\sin2\theta_{12}
\cos\theta_{13}\,\sin2\theta_{13}\, \sin\delta\,. 
\label{JCP2}
\end{equation}
%
Using eqs.\ (\ref{JCP}) and (\ref{eq:assumption}) we get \cite{HPR07}
\begin{equation}
J_{\rm CP} = - \frac{1}{8}\,\sin 2 \theta^{e}_{12} \, \sin 2 \hat \theta_{12} 
\, \sin 2 \hat \theta_{23} \, \sin \hat \theta_{23} \, \sin \phi\,. 
\label{JCP3}
\end{equation}
Thus, to leading order in $\sin \theta^{e}_{12} $ we have $\sin\delta = -  \sin \phi$. By  comparing the real part of $
U_{e1}^\ast \, U_{\mu 3}^\ast \, U_{e 3} \,
U_{\mu 1} $ in the two parameterisations we conclude that, at the same order, 
\begin{equation}
\delta = -  \phi\,.
\label{phidelta}
\end{equation}

\subsection{Relation between $\boldsymbol{\theta^e_{12}}$ and $\boldsymbol{\theta_{13}}$ in GUTs}

We would like now to study the possibility to generate a $\theta^e_{12}$ large enough to induce a  $\theta_{13}$ in the range indicated by recent experiments in the context of an SU(5) Grand Unified Theory (GUT). The unification assumption is powerful because it allows to relate the charged lepton and down quark Yukawa matrices $\lambda_E$ and $\lambda_D$. If all the Yukawa entries were generated by renormalizable operators and the MSSM Higgs fields were embedded in $\mathbf{5}$ and $\bar{ \mathbf{5}}$ representations only, we would have $\lambda^E_{ji} = \lambda^D_{ij}$, leading to wrong predictions for the
fermion mass ratios. In the general case one has instead $\lambda^E_{ji} = \alpha_{ij} \lambda^D_{ij}$. The coefficients $\alpha_{ij}$ depend on the operators from which the Yukawa entries arise. Such values can be constrained to belong to a finite set of rational numbers at the price of assuming that each Yukawa entry comes at least dominantly from a single renormalizable or non-renormalizable SU(5) operator\footnote{This could not be the case, for example, if SU(5) is embedded in SO(10) or a larger unified group.}. In this case, the possible values of the $\alpha_{ij}$ coefficients are listed in Table~\ref{Tab:SU5Relations}, see also \cite{Antusch:2009gu}. 

\begin{table}
\centering
\begin{tabular}{cc}\toprule
Operator Dimension & 	$\alpha_{ij}$ 	\\ 
\midrule 4	&	1	\\
	&	-3	\\
\midrule 5	&	-1/2	\\
	&	1	\\
	&	$\pm$3/2	\\
	&	-3	\\
	&	9/2	\\
	&	6	\\
	&	9	\\
	&	-18	\\
\bottomrule
\end{tabular}
\caption{
Summary of possible SU(5) predictions for the coefficients $\alpha_{ij}$. Numbers  are taken from \cite{Antusch:2009gu}, where also the corresponding operators are listed.
\label{Tab:SU5Relations}}
\end{table}

The $\theta^e_{12}$ angle is obtained from the diagonalization of the 12 block of the charged lepton Yukawa matrix after the 23 block has been diagonalized. Let us denote such 12 blocks in the charged lepton and down quark sectors (in the RL convention in which the Yukawa interactions are written with the left-handed fields on the right) as
\begin{equation}
\label{eq:y12}
\hat \lambda^D_{[12]} = 
\begin{pmatrix}
a & b' \\ b & c
\end{pmatrix}
\qquad
\hat \lambda^E_{[12]} = 
\begin{pmatrix}
\alpha a & \beta b \\ \beta' b' & \gamma c
\end{pmatrix} .
\end{equation}
In the following we will assume that the entries in \eq{y12} can be approximated with the corresponding entries of $\lambda^{E,D}$, 
in which case the coefficients $\alpha$, $\beta$, $\beta'$, $\gamma$ are still bound to take one of the values in Table~\ref{Tab:SU5Relations} (the rotation used to diagonalize the 23 sector can have a sizeable effect on the coefficient $\gamma$ and, if the charged lepton contribution to $\theta_{23}$ from $U^e_{23}$ is sizeable, on the coefficient $\beta$).

We would like to determine the values of the coefficients $\alpha$, $\beta$, $\beta'$, $\gamma$ allowed by data, and in particular capable to account for the indication for a sizeable $\theta_{13}$. Not all the values of the coefficients are allowed, in principle. The observables to be described are in fact
\begin{equation}
\label{eq:data}
\theta_{13}, \quad
|V_{us}|, \quad
\frac{m_e}{m_\mu}, \quad
\frac{m_d}{m_s}, \quad
\frac{m_\mu}{m_s} ,
\end{equation}
and for given $\alpha$, $\beta$, $\beta'$, $\gamma$, the five experimental inputs above depend on the four real variables $|b/c|$, $|b'/c|$, $|a/c|$, $\omega$, where the phase $\omega$ is defined by $ac(bb')^* = e^{i\omega}|acbb'|$. The explicit dependence is given by the following relations \globallabel{eq:relations}
\begin{align}
\tan\theta^e_{12} &= 
\left|
\frac{\beta' b'}{\gamma c}\left(
1-\fracwithdelims{|}{|}{\beta b}{\gamma c}^2
\right) + 
\frac{\beta b^*}{\gamma c^*}\frac{\alpha a}{\gamma c}
\right| \mytag \\
|V_{us}| &= \left|
\frac{b}{c}\left(
1-\fracwithdelims{|}{|}{b'}{c}^2 -\frac{1}{2}\fracwithdelims{|}{|}{b}{c}^2
\right) + 
\frac{b'^*}{c^*}\frac{a}{c}
\right| \pm\Delta \mytag \\
\frac{m_e}{m_\mu} &= \left|
\frac{\alpha}{\gamma} \frac{a}{c} - \frac{\beta\beta'}{\gamma^2} \frac{bb'}{c^2}
\right|
\left(
1- \frac{\beta^2|b|^2+\beta'^2|b'|^2}{\gamma^2|c|^2}
\right) \mytag \\
\frac{m_d}{m_s} &= \left|
\frac{a}{c} - \frac{bb'}{c^2}
\right|
\left(
1- \frac{|b|^2+|b'|^2}{|c|^2}
\right) \mytag \\ \label{eq:mmuOms}
\frac{m_\mu}{m_s} &= | \gamma |
\left(
1 +  \frac{(\beta^2-\gamma^2)|b|^2+(\beta'^2-\gamma^2)|b'|^2}{2|c|^2\gamma^2} 
\right) , \mytag 
\end{align}
where $\Delta$ takes into account the possibility of a model-dependent contribution to $|V_{us}|$ from the up quark sector and is assumed to be in the range $|\Delta| < \sqrt{m_u/m_c} \approx \text{0.045}$. The experimental inputs used for the quantities on the LHS are listed in Table~\ref{tab:inputs}. The relations above are approximated and are accurate up to corrections of order $\lambda^4$, if $|b/c| \lesssim |b'/c| \lesssim \lambda$, $|a/c| \lesssim \lambda^2$. 

Besides the general case in \eq{y12}, we will also consider the case in which $a=0$ and the symmetric case in which $|\lambda^D_{12}| = |\lambda^D_{21}|$ and $|\lambda^E_{12}| = |\lambda^E_{21}|$, as they arise in many models of fermion masses. Note that the symmetry condition implies $b= \pm b'$ and $\beta = \beta'$.

\begin{table}
\centering
\begin{tabular}{ccc} \toprule
Input Parameter & Value & Assumed error distribution \\ \midrule
$m_e/m_\mu$ & $ (4.7362 - 4.7369) \times 10^{-3}$ \cite{Xing:2007fb} & Uniform \\
$m_\mu/m_s$ & $2.48 - 7.73$ \cite{Antusch:2009gu} & Uniform \\
$m_s/m_d$ & $18.9 \pm 0.8$ \cite{Leutwyler:2000hx} & Gaussian \\
$|V_{us}|$ & $0.2252 \pm 0.0009$ \cite{PDG} & Used uniform in $|V_{us}| \pm \Delta$ \\
$\sin \theta_{13}$ & $0.089 - 0.190$ $(2\sigma)$ \cite{Fogli:2011qn} & Approximated with a Gaussian \\
\bottomrule
\end{tabular}
\caption{List of input parameters used in our analysis. \label{tab:inputs}}
\end{table}

\subsection{Procedure}

Before we come to the results we briefly discuss the procedure we implemented.
Since only the ratios $|a/c|$, $|b/c|$, and $|b'/c|$ enter when computing the experimental inputs, we have set in our forthcoming numerical analysis $|c| = 1$. We can also always perform a phase redefinition of the fields such that all the remaining coefficients are real and positive and the only physical phase is in $a$, so that $a = \text{exp} (\text{i} \, \omega)$.

For each possible combination of CG coefficients we diagonalized exactly both the mass matrices, using the expressions for the observables in eq.~\eqref{eq:data} in terms of $a,b,b',\omega$, of which the relations in eq.~\eqref{eq:relations} are the expansion at NLO. Then we determined numerically a solution for these parameters such that all the experimental inputs are satisfied.
We extracted the values for these inputs randomly following the distributions given in Table~\ref{tab:inputs}.
We repeated this procedure until one solution is found. If, after a large number of attempts, no solution is found, we discard this combination of CG coefficients.
For the viable CG coefficients we obtained by this procedure a distribution for $\theta_{13}$, from which we computed the mean value and the standard deviation.
To obtain $\sin \theta_{13}$ from $\sin \theta_{12}^e$ we have assumed that $\theta_{23}$ in the neutrino sector is maximal for simplicity. Given the uncertainties on the other input variables, this is a good approximation.

Note that eq.~\eqref{eq:mmuOms} fixes $\gamma$ to lie in the range of the observed $m_\mu/m_s$. Therefore we used this equation only to reduce the possible values of $\gamma$ to $-3$, 9/2 and 6, cf.\ Table~\ref{tab:inputs}. The GUT scale ratio $m_\mu/m_s$ depends strongly on low energy SUSY threshold corrections \cite{SUSYthresholds} and in principle one can use them to push this ratio to more extreme values, but in simple SUSY breaking scenarios these are the only plausible values \cite{Antusch:2009gu}.

\section{Results}

\begin{figure}
 \centering
 \includegraphics[scale=0.6]{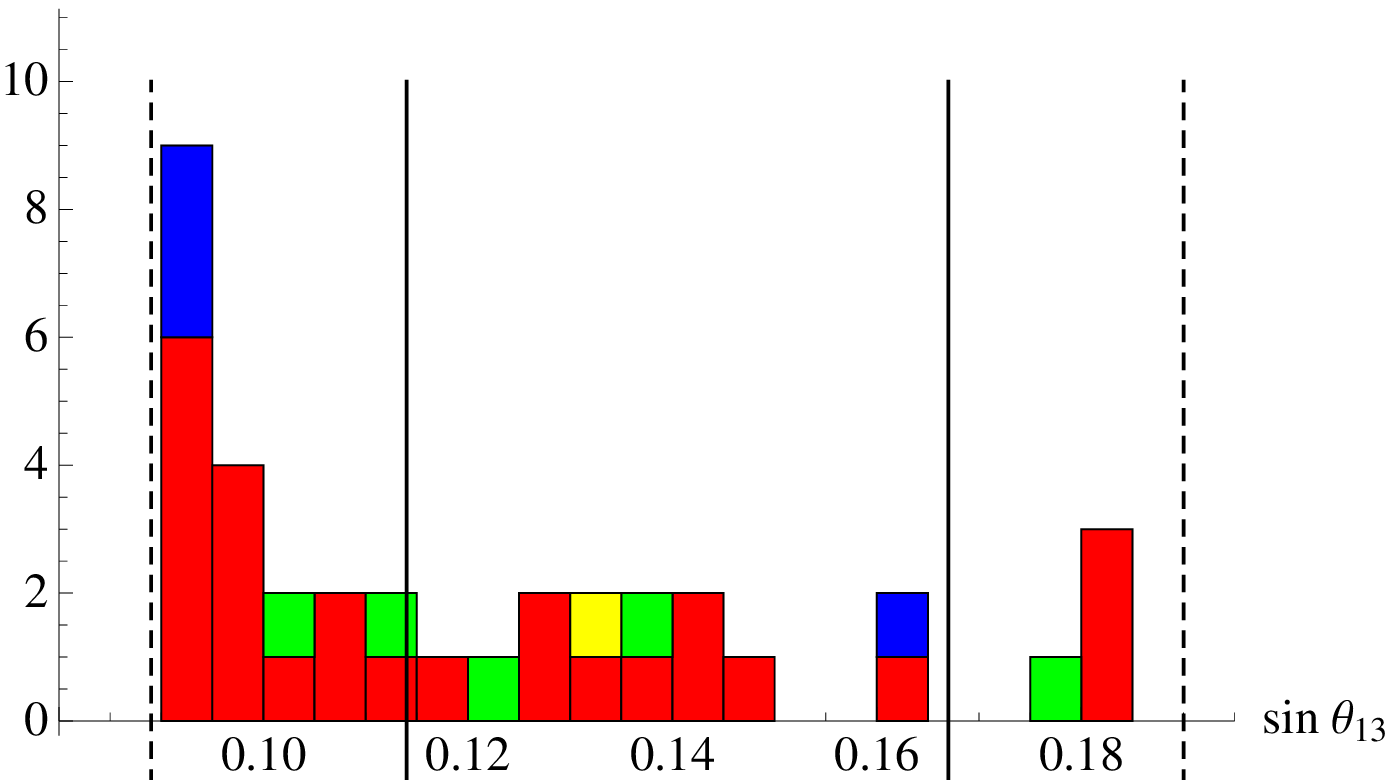}
 \caption{Graphical presentation of the results obtained. The vertical straight (dashed) lines denote the  
1$\sigma$ (2$\sigma$) allowed ranges of $\sin \theta_{13}$ taken from \cite{Fogli:2011qn} (see also Table~\ref{tab:experimental_PMNS_angles}). The yellow, green, blue, red bins correspond to the results from Table~\ref{tab:CG_gen_case_ren}, \ref{tab:CG_gen_case_adj}, \ref{tab:CG_gen_case_zero}, \ref{tab:CG_symm_case}. \label{fig:Histogram}}
\end{figure}

In the most general case it is easy to obtain values of the CG coefficients leading to a value of $\sin \theta_{13}$ compatible with the recent fits in \cite{Fogli:2011qn}. In fact we find several hundred possible combinations. We therefore 
restrict ourselves in the following to some well motivated cases. We give a graphical summary of the results at the end of this section in Fig.~\ref{fig:Histogram}.

\subsection{Results for Renormalizable Operators Only}

\begin{table}
\centering
\begin{tabular}{lcc} \toprule
$\{ \alpha, \beta, \beta', \gamma \}$ & $ \{ a, b, b', \omega \}$ & $\sin \theta_{13}$  \\ \midrule
$\{ -3, 1, -3, -3 \}$     &  $\{ 0.0151, 0.220, 0.189, -2.81 \} $ & $ 0.130 \pm 0.013 $ \\ \bottomrule
\end{tabular}
\caption{Possible CG coefficients with Yukawa couplings coming only from renormalizable operators. We also show typical values for the entries of $\hat{\lambda}_D$, where $c$ is normalised to one, and we give the prediction for $\sin \theta_{13}$ inlcuding its $1\sigma$ standard deviation. \label{tab:CG_gen_case_ren}}
\end{table}

We start our discussion with the case in which the Yukawa couplings come only from renormalizable operators. This case is very restrictive as there are only two possible CG coefficients, which are $\alpha_{ij} = 1$, if the Higgs sits in a $\mathbf{\bar{5}}$ of SU(5), and $\alpha_{ij} = -3$, if the Higgs sits in a $\mathbf{\overline{45}}$ of SU(5) \cite{Georgi:1979df}.

There is only one combination which is in agreement with the experimental data. It is shown in Table \ref{tab:CG_gen_case_ren}, where we give in addition typical values for the entries of $\hat{\lambda}_D$ and the prediction for $\sin \theta_{13}$.

\subsection{Results without Representations larger than the Adjoint}

\begin{table}
\centering
\begin{tabular}{lcc} \toprule
$\{ \alpha, \beta, \beta', \gamma \}$ & $ \{ a, b, b', \omega \}$ & $\sin \theta_{13}$  \\ \midrule
$\{ 1, -3/2, -3/2, 6 \}$     &  $\{ 0.0899, 0.246, 0.679, 0.145 \} $ & $ 0.114 \pm 0.014 $ \\
$\{ 1, -3/2, 6, 6 \}$     &  $\{ 0.0286, 0.212,  0.153, -2.57 \} $ & $ 0.103\pm 0.008 $ \\
$\{ -3/2, 1, 6, 6 \}$     &  $\{ 0.0224, 0.217, 0.186, -2.34 \} $ & $ 0.122 \pm 0.015 $ \\
$\{ 6, 1, 6, 6 \}$     &  $\{ 0.0155, 0.281, 0.259, 0.278 \} $ & $ 0.175 \pm 0.009 $ \\
$\{ 6, -3/2, 6, 6 \}$     &  $\{ 0.0134, 0.247, 0.184, -2.77 \} $ & $ 0.137 \pm 0.014 $ \\ \bottomrule
\end{tabular}
\caption{Possible CG coefficients with Higgs fields in representations not larger than the adjoint. We also show typical values for the entries of $\hat{\lambda}_D$, where $c$ is normalised to one, and we give the prediction for $\sin \theta_{13}$ inlcuding its $1\sigma$ standard deviation.  \label{tab:CG_gen_case_adj}}
\end{table}

The next case we consider is the one in which the Yukawa couplings are generated by a dimension five operator, with all fields sitting in a representation not larger than the adjoint. This concerns also the messenger sector of a possible UV completion. Especially the Georgi-Jarlskog factor of $-3$ \cite{Georgi:1979df} is here not possible anymore. There are only three $\alpha_{ij}$ left, which are $1$, $-3/2$, and $6$, giving five valid combinations as listed in Table \ref{tab:CG_gen_case_adj}, where we give again typical values for the parameters and the predictions for $\sin \theta_{13}$, including its standard deviation.

It is interesting to note that this possibility, as well as the last possibility, can be ruled out not only by a precise measurement of the leptonic mixing parameters, but also by a measurement of the SUSY spectrum.
A CMSSM like spectrum with a positive $\mu$ parameter prefers a ratio $m_\mu/m_s$ in the region of $4.5-6$ \cite{Antusch:2008tf, Antusch:2009gu}, ruling out the special case mentioned before. To get a small ratio $m_\mu/m_s \approx 3$, a spectrum more similar to an AMSB like scenario is preferred, in which the sign of the QCD part of the SUSY threshold corrections is flipped compared to the CMSSM with $\mu > 0$.

\subsection{Results for $\boldsymbol{a=0}$}

\begin{table}
\centering
\begin{tabular}{lcc} \toprule
$\{ \beta, \beta', \gamma \}$ & $ \{ b, b' \}$ & $\sin \theta_{13}$  \\ \midrule
$\{ -1/2, -3/2, -3 \}$     &  $\{ 0.217, 0.267 \} $ & $ 0.094 \pm 0.003 $ \\
$\{ -1/2, 3/2, -3 \}$     &  $\{ 0.216, 0.268 \} $ & $ 0.094 \pm 0.003 $\\
$\{ -1/2, 6, 6 \}$     &  $\{ 0.251, 0.240 \} $ & $ 0.164 \pm 0.013 $ \\
$\{ 1, -3, 6 \}$     &  $\{ 0.212, 0.273 \} $ & $ 0.094 \pm 0.003 $ \\ \bottomrule  
\end{tabular}
\caption{Possible Clebsch Gordan coefficients with a texture zero in the 11 element, $a=0$. We also show typical values for the entries of $\hat{\lambda}_D$, where $c$ is normalised to one, and we give the prediction for $\sin \theta_{13}$ inlcuding its $1\sigma$ standard deviation. \label{tab:CG_gen_case_zero}}
\end{table}

The next scenario we discuss is a scenario, where we have a texture zero in the 11 element, $a=0$. This can be motivated by having a flavon vacuum alignment, which has a zero in this position or having a Froggat-Nielsen mechanism at work, which puts there a zero or suppresses this element very strongly.
For the CG coefficients we take all the possible values in Table \ref{Tab:SU5Relations}. In this case we end up with four possible combinations, which are listed in Table \ref{tab:CG_gen_case_zero}. Note that in this case there are no physical phases.

\subsection{Results for Symmetric Mass Matrices}
\begin{table}
	\centering
	\begin{tabular}[]{l c  c} \toprule
	$\{ \alpha, \beta, \gamma \}$ & $\{ a,b,\omega \}$ & $\sin \theta_{13}$ \\ \midrule
  $\{-1/2, -3/2, -3\} $& $\{ 0.122, 0.259, -0.183\}$ &$ 0.0903 \pm 0.0008 $ \\
 $ \{-1/2, 3/2, -3\} $& $\{ 0.125, 0.255, -0.0985\}$ &$ 0.0903 \pm 0.0008 $ \\
$\{-3/2, -3, -3\} $& $\{ 0.115, 0.233, -0.0736\}$ &$ 0.164 \pm 0.007 $ \\
  $\{-3, -3/2, -3\} $& $\{ 0.0205, 0.284, 0.243\}$ &$ 0.098 \pm 0.002 $ \\
  $\{-3, 3/2, -3\} $& $\{ 0.0143, 0.268, 0.201\}$ &$ 0.098 \pm 0.002 $ \\
  $\{6, -3, -3\} $& $\{ 0.0186, 0.205, -3.08\}$ &$ 0.139 \pm 0.001 $ \\
  $\{9, -3, -3\} $& $\{ 0.0142, 0.212, -3.04\}$ &$ 0.144 \pm 0.003 $\\
  $\{-18, -3/2, -3\} $& $\{ 0.0028, 0.257, -0.294\}$ &$ 0.0901 \pm 0.0008 $\\
  $\{-18, 3/2, -3\} $& $\{ 0.0033, 0.255, -0.187\}$ &$ 0.0900 \pm 0.0007 $\\
  $\{-18, -3, -3\} $& $\{ 0.0120, 0.268, -0.0757\}$ &$ 0.183 \pm 0.004 $\\ \midrule
 $ \{1, -3, 9/2\} $& $\{ 0.115, 0.234, 0.195\}$ &$ 0.105 \pm 0.009 $\\
  $\{3/2, 9/2, 9/2\} $& $\{ 0.093, 0.186, 0.107\}$ &$ 0.128 \pm 0.003 $\\
  $\{6, -3, 9/2\} $& $\{ 0.0254, 0.289, 0.108\}$ &$ 0.1325 \pm 0.0009 $\\
  $\{9, -3, 9/2\} $& $\{ 0.0155, 0.275, 0.137\}$ &$ 0.129 \pm 0.003 $\\
  $\{-18, -3, 9/2\} $& $\{ 0.0057, 0.240, 2.97\}$ &$ 0.107 \pm 0.002 $\\
  $\{-18, 9/2, 9/2\} $& $\{ 0.0117, 0.209, -3.05\}$ &$ 0.149 \pm 0.003 $\\
  $\{-18, 6, 9/2\} $& $\{ 0.0183, 0.258, -3.10\}$ &$ 0.184 \pm 0.002 $\\ \midrule
  $\{1, -3, 6\} $& $\{ 0.127, 0.258, -0.114\}$ &$ 0.0903 \pm 0.0008$ \\
  $\{1, 9/2, 6\} $& $\{ 0.0888, 0.181, 0.200\}$ &$ 0.097 \pm 0.002 $\\
  $\{3/2, 9/2, 6\} $& $\{ 0.111, 0.225, 0.177\}$ &$ 0.11 \pm 0.01 $\\
  $\{9/2, -3, 6\} $& $\{ 0.0200, 0.280, 0.105\}$ &$ 0.100 \pm 0.001 $\\
  $\{9/2, 9, 6\} $& $\{ 0.0918, 0.179, 0.060\}$ &$ 0.183 \pm 0.002$ \\
  $\{6, -3, 6\} $& $\{ 0.0182, 0.280, 0.249\}$ &$ 0.098 \pm 0.002 $\\
  $\{9, -3, 6\} $& $\{ 0.0108, 0.263, 0.286\}$ &$ 0.094 \pm 0.003$ \\
  $\{-18, 9/2, 6\} $& $\{ 0.0094, 0.221, -2.96\}$ &$ 0.116 \pm 0.002 $\\
  $\{-18, 6, 6\} $& $\{ 0.0133, 0.211, -3.08\}$ &$ 0.143 \pm 0.003 $\\ \bottomrule
	\end{tabular}
	\caption{Possible Clebsch-Gordan coefficients with a symmetric mass matrix and the resulting prediction for $\sin \theta_{13}$.   \label{tab:CG_symm_case} }	
\end{table}

In the (anti-)symmetric case $|\lambda^D_{12}| = |\lambda^D_{21}|$ and $|\lambda^E_{12}| = |\lambda^E_{21}|$, which implies $b= \pm b'$ and $\beta = \beta'$, we find 26 possible combinations listed in Table~\ref{tab:CG_symm_case}. Such a mass matrix is generated, if the 12 and the 21 entries are coming from the same operator.
Note that by choosing the unphysical phases appropriately we can always make $b = b'$.

This case cannot be combined with any other case. If we restrict ourselves to certain operators or choose $a=0$, no combination remains viable.

\section{Possible Implications for Dirac CP Violation in the Lepton Sector}

The results of the T2K and MINOS 
experiments \cite{Abe:2011sj,MINOS240611}
and of the global analysis 
of the neutrino oscillation data \cite{Fogli:2011qn} 
have important implications for the Dirac CP violation 
in the lepton sector when the unitary mixing 
matrix $U^\dagger_\nu$, originating from the 
diagonalisation of the neutrino mass matrix, is 
(up to CP violating 
diagonal phase matrices) of 
BM or TBM form, 
while that originating from the 
diagonalisation of the charged lepton mass matrix 
can be approximated as $U_e =  R_{12}(\theta^e_{12})$
(see eq.\ (\ref{eq:assumption})). 
We recall that if $U^\dagger_\nu$ coincides with 
the BM mixing matrix one has:
\begin{equation}
 \sin^2 \theta^\nu_{12} = \frac{1}{2} \;, \quad \sin^2 \theta^\nu_{23} = \frac{1}{2} \;, \quad  \sin^2 \theta^\nu_{13} = 0 \;. 
\end{equation}
%
In the case of  tri-bimaximal mixing form of $U^\dagger_{\nu}$, 
\begin{equation}
 \sin^2 \theta^\nu_{12} = \frac{1}{3} \;, \quad \sin^2 \theta^\nu_{23} = \frac{1}{2} \;, \quad  \sin^2 \theta^\nu_{13} = 0 \;. 
\end{equation}
%
While in both cases $\theta^{\nu}_{23}$ coincides, or is close to, 
the experimentally determined best fit value 
of the atmospheric neutrino mixing angle
$\theta_{23}$,
in the case of bimaximal mixing form 
of  $U^\dagger_\nu$ a relatively 
large correction to $\theta^{\nu}_{12}$ is needed 
to get a value of the solar neutrino mixing angle 
$\theta_{12}$ compatible with that determined from the data. 
And in both cases a non-zero $\theta_{13}$, having a value
in the range 
$\sin^2\theta_{13} = 0.021~(0.025) \pm 0.007$, 
see also eq.~(\ref{th13Fogli}),
suggested by the current data, has to be generated.

  If $U^\dagger_\nu$ has bimaximal form,
the angle $\theta^{\nu}_{12} = \pi/4$ 
is corrected by the charged 
lepton mixing  as follows:
\begin{equation}
\sin^2 \theta_{12} 
\simeq \frac{1}{2} \frac{1 + 
\frac{1}{\sqrt{2}} \cos \phi \sin 2 \theta^e_{12}
- \frac{1}{2} \sin^2 \theta^e_{12}}
{1 - \frac{1}{2} \sin^2 \theta^e_{12} }\,,~~~\text{(BM)}\,,
\label{th12BM1}
\end{equation}
%
where we have neglected the possible contributions from the 
charged lepton mixing angles $\theta_{13}^e$ and $\theta_{23}^e$~
\footnote{These contributions are given in \cite{FPR04,HPR07}.}. 
To leading order in $\sin \theta^e_{12}$ we get 
\cite{FPR04,Romanino:2004ww,HPR07,alta}:
\begin{equation}
\sin^2 \theta_{12} \simeq 
\frac{1}{2} + \frac{1}{\sqrt{2}}\cos \phi\, \sin \theta^e_{12}\,
\simeq \frac{1}{2} + \cos \delta\, \sin \theta_{13}\,,
~~~\text{(BM)}\,,
\label{th12BM2}
\end{equation}
%
where we have used the relation 
$\cos \phi = \cos \delta$ and the fact that 
in the approximation employed we have:
\begin{eqnarray}
\sin \theta_{13} \simeq \frac{1}{\sqrt{2}} \sin \theta^e_{12} > 0\,.
\label{th13BMTBM}
\end{eqnarray}
%
The sign of the second term in the r.h.s.\ of 
the equations in (\ref{th12BM2}) 
is important 
\footnote{
We note that in \cite{FPR04,HPR07,alta}
a somewhat different parametrisation of the PMNS matrix 
was used (namely, $U = U^\dagger_e U_\nu$ and 
$U = R^T_{12}(\theta^e_{12}) \Phi R_{23}(\hat \theta_{23}) 
R_{12}(\hat \theta_{12})Q$) and instead of the first 
relation in (\ref{th12BM2}), the relation 
$\sin^2 \theta_{12} \simeq 1/2 - (\cos \phi\, \sin \theta^e_{12})/\sqrt{2}$
was obtained. It is not difficult to show that 
in this case  we have $\cos\delta = - \cos \phi$
and thus one arrives at the same result 
for the relation between $\sin^2 \theta_{12}$, $\theta_{13}$ and $\delta$.
}
in view of the fact that, 
according to \cite{Fogli:2011qn}, 
the global neutrino oscillation data favors
a negative value of $\cos \delta \sin \theta_{13}$
over the positive value. 
We note that such a sign is in agreement 
with the one needed to reduce the bimaximal 
prediction for $\theta_{12}$ 
down to the experimentally allowed range.

In the case of the TBM form of  $U^\dagger_{\nu}$
we find under the same assumptions:
\begin{equation}
\sin^2 \theta_{12} \simeq 
\frac{1}{3} \frac{1 + \cos \phi \sin 2 \theta^e_{12}\,
}
{1 - \frac{1}{2} \sin^2 \theta^e_{12} }\;,~~~\text{(TBM)}.
\label{th12TBM1}
\end{equation}
%
To leading order in $\sin\theta^e_{12}$ we get:
\begin{equation}
 \sin^2 \theta_{12}\simeq \frac{1}{3} + \frac{2\sqrt{2}}{3}\cos \delta\, \sin\theta_{13}\,, 
~~~\text{(TBM)}\,,
\label{th12TBM2}
\end{equation}
%
where we have used again 
$\cos \phi = \cos \delta$ 
and eq.~(\ref{th13BMTBM}), 
which is valid also in this case.

 The angle $\theta^{\nu}_{23} = \pi/4$ gets the same 
correction in both cases of BM and TBM
$U^\dagger_{\nu}$:
\begin{eqnarray}
 \sin^2 \theta_{23} \simeq \frac{1}{2} 
\frac{\cos^2 \theta_{12}^e}{1 - \frac{1}{2} \sin^2 \theta_{12}^e } \;.
\label{th23BMTBM}
\end{eqnarray}
%
As it follows from the above expression, 
the leading correction is 
of order $\sin^2 \theta_{12}^e$ \cite{FPR04}.
\begin{table}
\centering
\begin{tabular}{l c c c} \toprule
			   & $\sin \theta_{13}$ & $\sin^2 \theta_{12}$  & $\sin^2 \theta_{23}$ \\ \midrule
	Best fit	   & 0.145		    & 		0.306    &    0.42\\
	$1 \sigma $ & 0.114 - 0.167 & 0.291 - 0.324 & 0.39 - 0.50 \\
	$2 \sigma $ & 0.089 - 0.190 & 0.275 - 0.342 & 0.36 - 0.60 \\
	$3 \sigma $ & 0.032 - 0.210 & 0.259 - 0.359 & 0.34 - 0.64 \\ \bottomrule
\end{tabular}
\caption{Results of the global fit of the PMNS mixing 
angles taken from \cite{Fogli:2011qn} and used in our analysis.
The results quoted were obtained using the 
``old'' reactor $\bar{\nu}_e$ fluxes 
(see \cite{Fogli:2011qn} for details). 
\label{tab:experimental_PMNS_angles}}
\end{table}

\begin{figure}
 \centering
 \includegraphics[scale=0.6]{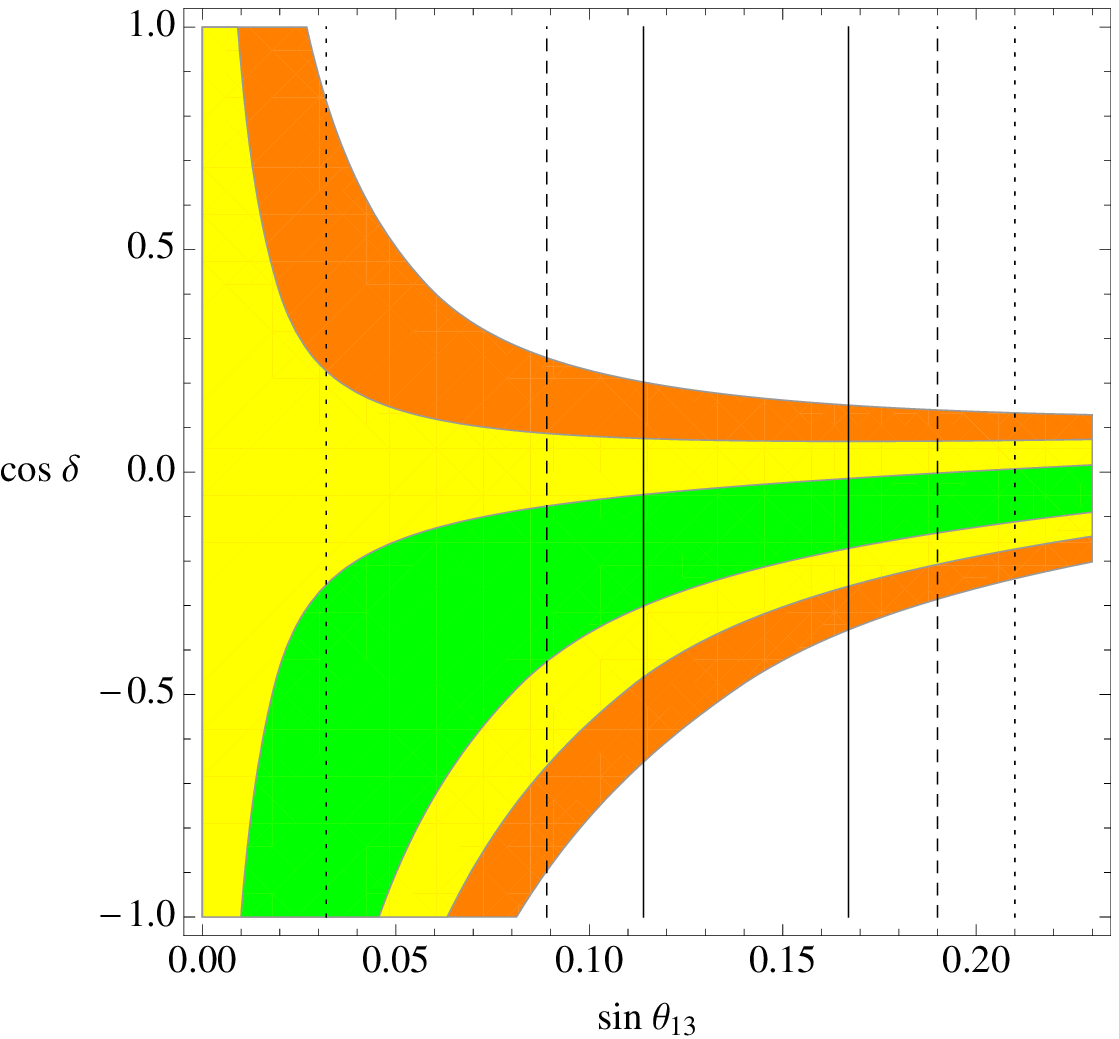} \hspace{0.2cm}
\includegraphics[scale=0.6]{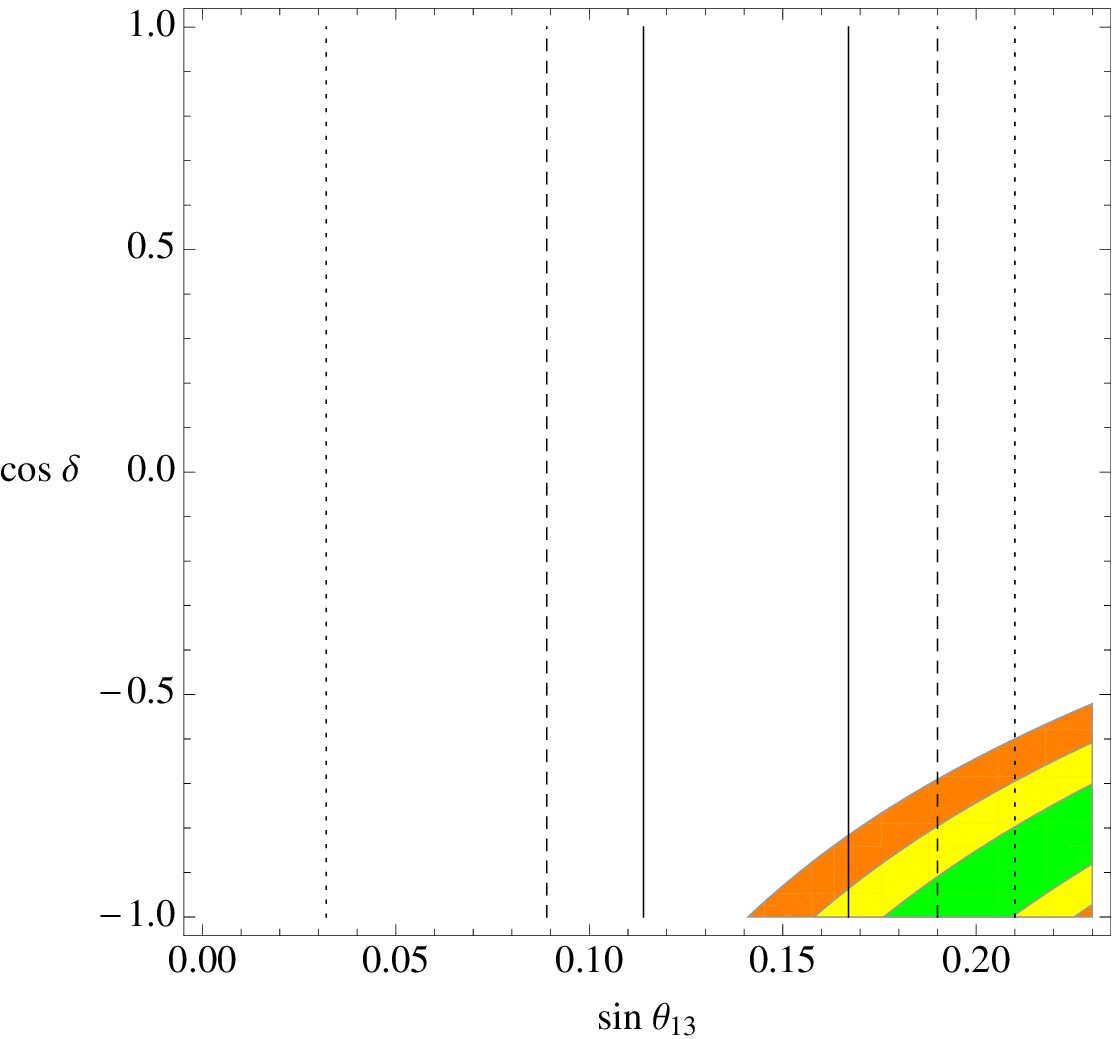}
 \caption{
The cosine of the Dirac CP phase $\delta$
as a function of $\sin \theta_{13}$ in the cases of 
tri-bimaximal (left panel) and bimaximal mixing (right panel) 
arising from the diagonalisation of the neutrino mass matrix,
eqs. (\ref{th12BM1}) and (\ref{th12TBM1}). 
The green, yellow, orange regions correspond to   
the 1, 2, 3$\sigma$ allowed ranges of $\sin^2 \theta_{12}$. 
In the bimaximal mixing case only the negative 
values of $\cos\delta$ are compatible 
with the data on $\sin^2\theta_{12}$. 
The vertical straight, dashed, dotted lines denote the  
1, 2, 3$\sigma$ allowed ranges of $\sin \theta_{13}$. 
The values of $\sin^2 \theta_{12}$ and  $\sin \theta_{13}$ 
are taken from \cite{Fogli:2011qn} (see text for details).
\label{fig:CP1}}
\end{figure}

  The values of  $\sin^2 \theta_{12}$, $\sin^2 \theta_{23}$ and
$\sin^2 \theta_{13}$, obtained in the global data analysis 
in \cite{Fogli:2011qn} are shown for convenience
in Table~\ref{tab:experimental_PMNS_angles}. 
In Fig.~\ref{fig:CP1} we present graphically the 
constraints on $\cos \delta$ implied by the data 
on $\sin^2 \theta_{12}$ and $\sin \theta_{13}$
in both cases of BM and TBM 
$U^\dagger_{\nu}$, using the relations~(\ref{th12BM1}, 
\ref{th12TBM1}), which are exact in $\theta^e_{12}$.

 A few comments are in order.
It follows from our analysis that in the case 
of bimaximal $U^\dagger_{\nu}$, the current $3\sigma$ ($2\sigma$)
experimentally allowed range for $\sin^2 \theta_{12}$
requires that  $\sin \theta_{13} \gtap 0.14~(0.16)$,
the minimal value being very close to  the best 
fit value found in  \cite{Fogli:2011qn}.
If future data on $\theta_{13}$ will show 
(taking into account all relevant 
uncertainties) that, e.g.\ 
$\sin \theta_{13} \ltap 0.10$,  
the simple case under discussion 
of  $U^\dagger_{\nu}$ having a BM mixing 
form and $U_e = R_{12}(\theta^e_{12})$
will be ruled out.
Further, using the $3\sigma$ ($2\sigma$) allowed ranges 
of both $\sin^2 \theta_{12}$ and $\sin \theta_{13}$
we find that $\cos\delta$ is constrained to lie
in the interval: $-1\leq  \cos \delta \ltap -0.60~(-0.79)$.
Thus, $\cos\delta = 0$, or $\delta = \pi/2$,
and therefore maximal CP violation 
in neutrino oscillations,
is ruled out in the  
scheme we are considering.

 We get very different constraints on 
$\cos\delta$ in the case of the TBM
form of  $U^\dagger_{\nu}$. If we use the 
$3\sigma$ allowed ranges of 
$\sin^2 \theta_{12}$ and $\sin \theta_{13}$ 
in the analysis, all possible values of 
$\cos\delta$ are allowed. The $2\sigma$ 
intervals of allowed values of 
$\sin^2 \theta_{12}$ and $\sin \theta_{13}$
require that
$-0.66\ltap  \cos \delta \ltap 0.09$. 
Thus, maximal CP violation, 
$\delta = \pi/2$, 
is allowed.

\begin{figure}
 \centering
 \includegraphics[scale=0.6]{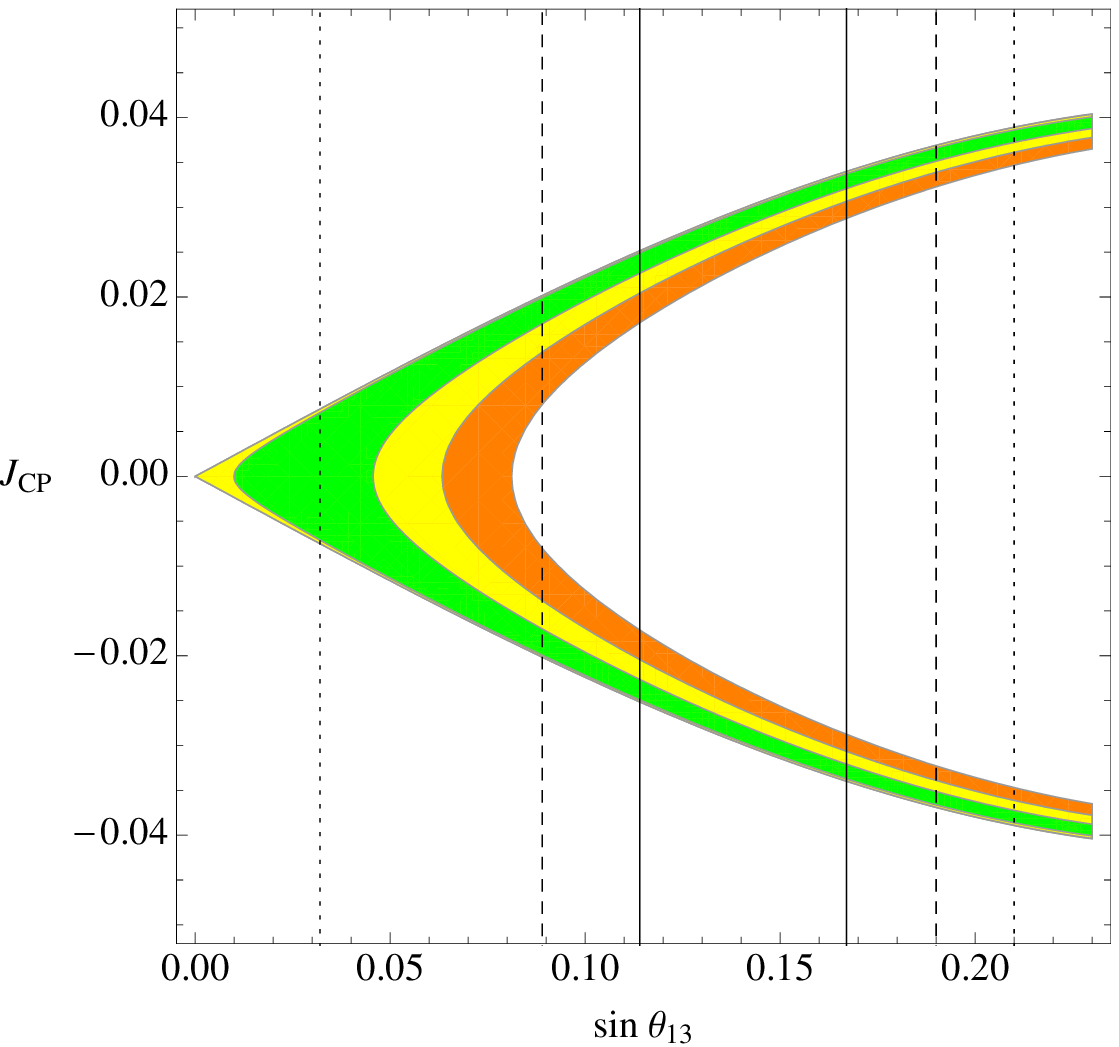} \hspace{0.2cm}
 \includegraphics[scale=0.6]{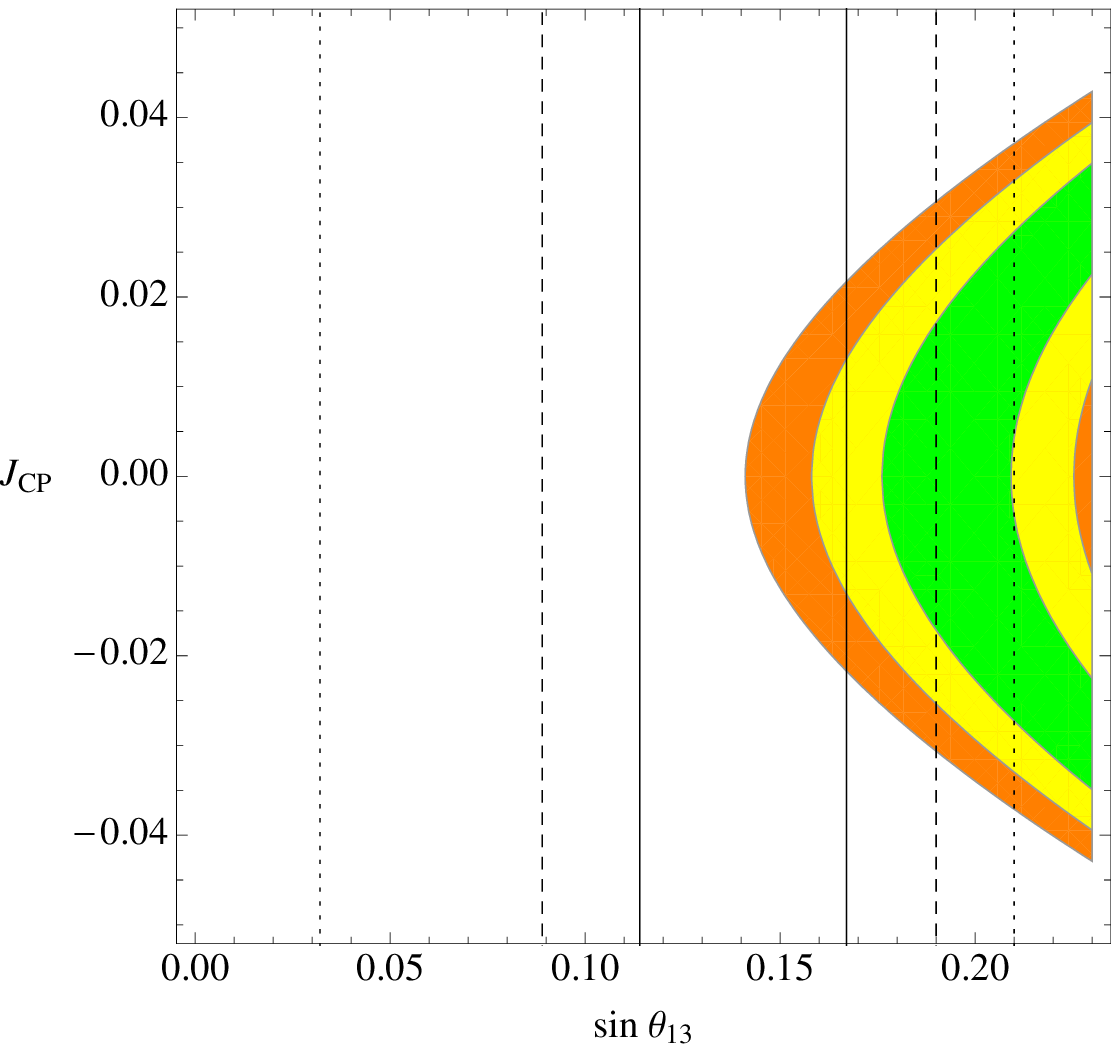}
 \caption{The rephasing invariant $J_{\text{CP}}$ 
as a function of $\sin \theta_{13}$
in the cases of tri-bimaximal (left) and bi-maximal (right) 
mixing from the neutrino sector. 
The green, yellow, orange regions correspond to the 1, 2, 3$\sigma$ 
allowed ranges of $\sin^2 \theta_{12}$. 
The vertical straight, dashed, dotted lines show  
the  1, 2, 3$\sigma$ allowed ranges of $\sin \theta_{13}$. 
The 1, 2, 3$\sigma$ allowed ranges of
$\sin^2 \theta_{12}$ and  $\sin \theta_{13}$  
are the same as in Fig. \ref{fig:CP1}.
\label{fig:CP2}}
\end{figure}

Using the constraints on $\delta$ derived above,
we have obtained also predictions for the magnitude of the 
rephasing invariant $J_{\text{CP}}$. 
The latter has the following form to
leading order in $\sin\theta^e_{12}$
in the cases of BM and TBM $U^\dagger_{\nu}$:
\begin{eqnarray}
 J_{\text{CP}}^\text{BM}~ &\simeq& 
\frac{1}{4} \sin \delta \sin \theta_{13}
\,,\\ 
J_{\text{CP}}^\text{TBM}&\simeq& \frac{1}{3\sqrt{2}} \sin \delta 
\sin \theta_{13}
\,.
\label{JCPTBM2}
\end{eqnarray}
%

The results of this analysis are presented 
in Fig.~\ref{fig:CP2}. As could be expected 
on  the basis of the results obtained for 
$\cos\delta$, the existing data on 
$\sin^2 \theta_{12}$ and $\sin \theta_{13}$ 
imply that in the case of BM form of
$U^\dagger_{\nu}$, $J_{\text{CP}}$ should lie in the interval
$|J_{\text{CP}}| \lesssim 0.037 (0.031)$, using
the $3\sigma$ ($2\sigma$) ranges, the value 
$J_{\text{CP}} = 0$ being allowed.
We get similar results for the maximal possible 
value of $|J_{\text{CP}}|$ if $U^\dagger_{\nu}$ exhibits 
TBM mixing form.
However, if we use the $2\sigma$ allowed ranges
of $\sin^2 \theta_{12}$ and $\sin \theta_{13}$, 
we find that $|J_{\text{CP}}|$ has to be non-zero and 
not smaller than  0.014: 
$0.014 \lesssim |J_{\text{CP}}| \lesssim 0.037$.
Values of $|J_{\text{CP}}| \gtap 0.01$ are potentially 
accessible by the accelerator experiments T2K, NO$\nu$A, etc.\ 
planning to search for CP violation effects in neutrino 
oscillations (see, e.g.\ \cite{Bernabeu:2010rz}).

\section{Summary and Conclusions}

In the present article we have considered 
the possibility that a sizeable value of the neutrino 
mixing angle $\theta_{13}$, compatible with the recent 
indications from the T2K, MINOS and 
the global neutrino oscillation data, arises 
primarily from 
the contribution of the charged 
lepton sector $U_e$ to the lepton mixing, 
$U = U_e\,U^\dagger_{\nu}$, where $U$ is the PMNS 
matrix and $U^\dagger_{\nu}$ arises from the 
diagonalisation of the neutrino mass matrix.
This scenario 
should necessarily be taking place
in the context of a wide variety of 
models predicting a relatively small 
contribution to $\theta_{13}$ from the neutrino sector, 
say $\theta_{13}\lesssim \mathcal{O}(\lambda^2)$, 
where $\lambda$ is the Cabibbo expansion parameter 
$\lambda\approx 0.23$. The analysis has been performed 
in the context of SU(5) supersymmetric models, which are 
independently well motivated and allow the charged 
lepton contribution to $\theta_{13}$ to be related to 
the mass and mixing observables in the quark sector. 

The naive relation $m_E = m_D^T$ between the charged lepton and 
down quark mass matrices is not compatible with the 
indication of a relatively large $\theta_{13}$. 
This is not surprising, as the quoted
relation is not compatible with the 
well known values of the charged fermion mass 
ratios either and 
it has to be corrected, possibly
by Clebsch-Gordan (CG) factors. 
This is 
particularly true in the case 
(relevant for us) of the 
charged fermions belonging to the first 
two families, whose small masses 
can arise from non-renormalizable 
operators that can well involve SU(5) 
breaking sources. Under motivated hypotheses 
for the values of such CG coefficients, 
we found all the combinations that 
are experimentally viable. 
The main results are summarized in 
Tables~\ref{tab:CG_gen_case_ren},~\ref{tab:CG_gen_case_adj},
~\ref{tab:CG_gen_case_zero}, and \ref{tab:CG_symm_case}. 
In order to describe the present data, one needs either a 
non-vanishing 11 entry in the down quark and charged lepton mass matrices, 
or CG coefficients from non-renormalizable 
or from SU(5) breaking operators 
in representations larger than the adjoint. 

We also studied the specific case in which the 
contribution from the neutrino sector to the lepton 
mixing, $U^\dagger_{\nu}$ in eq.\ (\ref{eq:U}), is,
up to the phase matrices $\Phi$ and $Q$, see, eq.\ \eqref{eq:assumption},
of the  bimaximal (BM) or tri-bimaximal 
(TBM) form,
while that from the charged lepton sector
is assumed to be of the form $U_e = R_{12}(\theta^e_{12})$.
Under these assumptions, $\Phi = \mathrm{diag}(1,e^{i \phi},1)$, 
$\phi$ being a CP violating phase, and $Q$ contains the two Majorana CP violating phases. 
In this case, it is possible to draw 
important conclusions on 
the leptonic (Dirac) CP-violating phase $\delta \approx - \phi$
and on the magnitude of the CP violation effects 
in neutrino oscillations. The latter, as is well known, 
is determined by the rephasing invariant $J_\text{CP}$. 
Our results are summarized 
in Figs.~\ref{fig:CP1},~\ref{fig:CP2}. 
We find that in the case 
of bimaximal $U^\dagger_{\nu}$, 
the current $3\sigma$ ($2\sigma$)
experimentally allowed range for $\sin^2 \theta_{12}$
requires that  $\sin \theta_{13} \gtap 0.14~(0.16)$.
If future data on $\theta_{13}$ will show 
(taking into account all relevant 
uncertainties) that $\sin \theta_{13}$ has a smaller value, 
e.g.\ $\sin \theta_{13} \ltap 0.10$,  
the simple case of  $U_{\nu}$ having a BM mixing 
form and $U_e = R_{12}(\theta^e_{12})$
will be ruled out.
Further, using the $3\sigma$ ($2\sigma$) allowed ranges 
of both $\sin^2 \theta_{12}$ and $\sin \theta_{13}$
we find that $\cos\delta$ is constrained to lie
in the interval: $-1\leq  \cos \delta \ltap -0.60~(-0.79)$.
Thus, $\cos\delta = 0$, or $\delta = \pi/2,3\pi/2$,
and therefore maximal CP violation 
in neutrino oscillations, is essentially ruled out
in the scheme we have considered.
If it will be confirmed experimentally that 
$\cos \delta\, \sin\theta_{13} < 0$, 
that would imply  
(in the standard parameterization of the PMNS matrix, 
in which $\sin\theta_{13} > 0$) that 
$127^\circ~(142^\circ) \ltap  \delta \ltap 233^\circ~(218^\circ)$.
Correspondingly, $J_{\text{CP}}$ should lie in the
$3\sigma$ ($2\sigma$) interval
$|J_{\text{CP}}| \lesssim 0.037~(0.031)$, the value 
$J_{\text{CP}} = 0$ being allowed.
However, if it will be experimentally established 
that $\cos \delta\, \sin\theta_{13} > 0$,
that would rule out the case of $U_e = R_{12}(\theta^e_{12})$
and $U^\dagger_{\nu}$ having a BM mixing form.

In the case of the TBM
form of  $U^\dagger_{\nu}$
and $U_e = R_{12}(\theta^e_{12})$,
all possible values of 
$\cos\delta$ are allowed
if one uses the $3\sigma$ allowed ranges of 
$\sin^2 \theta_{12}$ and $\sin \theta_{13}$ 
in the analysis. 
The $2\sigma$ intervals of allowed values of 
$\sin^2 \theta_{12}$ and $\sin \theta_{13}$
require that $-0.66\ltap  \cos \delta \ltap 0.09$,
or   $85^\circ\ltap  \delta \ltap 131^\circ$.
Thus, maximal CP violation, $\delta = \pi/2$, 
in  both cases is allowed.
We find using  $2\sigma$ allowed ranges
of $\sin^2 \theta_{12}$ and $\sin \theta_{13}$, 
that $|J_{\text{CP}}|$ has to be non-zero and 
not smaller than  0.014: 
$0.014 \lesssim |J_{\text{CP}}| \lesssim 0.037$.
This interval falls in the range of the potential sensitivity 
of the T2K and the future neutrino neutrino oscillation 
experiments (NO$\nu$A, etc.), designed to search for 
CP violation effects in neutrino oscillations.

\section*{Acknowledgements}
This work was supported in part by the INFN program 
on ``Astroparticle Physics'', by the Italian MIUR program 
on ``Neutrinos, Dark Matter and  Dark Energy in the Era of LHC'' and 
by the World Premier International 
Research Center Initiative (WPI Initiative), 
MEXT, Japan  (S.T.P.), by the EU Marie Curie ITN ``UNILHC'' 
(PITN-GA-2009-23792), and by the ERC Advanced Grant no. 267985 ``DaMESyFla''.

\vspace{0.6cm}
{\bf Note Added.}
During the finalising stage of the work on the present article, 
Ref.\ \cite{Antusch:2011qg} appeared, where the authors 
follow a similar approach in generating a relatively 
large $\theta_{13}$.
Our work goes beyond the analysis performed in 
\cite{Antusch:2011qg} in a number of aspects.
In \cite{Antusch:2011qg} 
only the case $a=0$ is discussed,
while we consider the more general case
of $a\neq 0$.
The authors of \cite{Antusch:2011qg}  
use only leading order terms (without error estimates) 
in the expressions for the observables in terms of 
the parameters of the down quark and charged lepton 
mass matrices (i.e., in  eqs. (12a) - (12e))  
and they do not utilise the 
rather precisely known value of the ratio  $m_s/m_d$ 
(see Table~\ref{tab:inputs}).
This rules out some of the possibilities 
considered in  \cite{Antusch:2011qg} to be viable.
Finally, the  authors of \cite{Antusch:2011qg} 
draw only  qualitatively conclusions about the 
value Dirac CP violating phase $\delta$ 
in the case of (tri-)bimaximal mixing 
arising from the neutrino sector. 
We perform a quantitative analysis and determine the 
allowed ranges (at 1,2 and 3 $\sigma$) of values of 
$\cos\delta$ and of the rephasing invariant $J_\text{CP}$ 
in the two cases (Figs.  \ref{fig:CP1} and  \ref{fig:CP2}).

\end{document}